# MgB$_2$ single crystals substituted with Li and with Li-C: Structural and superconducting properties

[a]J. Karpinski, [a]N. Zhigadlo, [a]S. Katrych, [a]B. Batlogg, [b,a]M. Tortello, [c,a]K. Rogacki, [d,a]R. Puzniak

[a]*Laboratory for Solid State Physics, ETH, 8093 Zurich, Switzerland*

[b]*Dipartimento di Fisica and CNISM, Politecnico di Torino, 10129 Torino, Italy*

[c]*Institute of Low Temperature and Structure Research, Polish Academy of Sciences, 50-950 Wroclaw, P.O. Box 1410, Poland*

[d]*Institute of Physics, Polish Academy of Sciences, Aleja Lotnikow 32/46 02-668 Warsaw, Poland*

Abstract

The effect of Li substitution for Mg and of Li-C co-substitution on the superconducting properties and crystal structure of MgB$_2$ single crystals has been investigated. It has been found that hole doping with Li decreases the superconducting transition temperature $T_c$, but at a slower rate than electron doping with C or Al. $T_c$ of MgB$_2$ crystals with simultaneously substituted Li for Mg and C for B decreases more than in the case where C is substituted alone. This means that holes introduced by Li cannot counterbalance the effect of decrease of $T_c$ caused by introduction of electrons coming from C. The possible reason of it can be that holes coming from Li occupy the π band while electrons coming from C fill the σ band. The temperature dependences of the upper critical field $H_{c2}$ for Al and Li substituted crystals with the same $T_c$ show a similar d$H_{c2}$/d$T$ slope at $T_c$ and a similar $H_{c2}(T)$ behavior, despite of much different substitution level. This indicates that the mechanism controlling $H_{c2}$ and $T_c$ is similar in both hole and electron doped crystals. Electrical transport measurements show an increase of resistivity both in Li substituted crystals and in Li and C co-substituted crystals. This indicates enhanced scattering due to defects introduced by substitutions including distortion of the lattice. The observed behavior can be explained as a result of two effects, influencing both $T_c$ and $H_{c2}$. The first one is doping related to the changes in the carrier concentration, which may lead to the decrease or to the increase of $T_c$. The second one is related to the introduction of new scattering centers leading to the modification of the interband and/or intraband scattering and therefore, to changes in the superconducting gaps and to the reduction of $T_c$.

# I. INTRODUCTION

After several years of intensive investigations, the superconducting and normal state properties of pure $MgB_2$ are now well explored by experiment and explained by theory. $MgB_2$ is a two-band two-gap superconductor with several anomalous properties originating from the existence of two separate sheets of the Fermi surface: first one quasi 2D ($\sigma$ band) and a second one quasi 3D ($\pi$ band) [1-5]. This electronic structure and strong electron-phonon coupling predominantly on the $\sigma$ sheet lead to the high critical temperature $T_c$ of 39 K, and to a pronounced temperature and field dependent anisotropy [6] of electronic properties, particularly in the superconducting state. The evolution of two gaps as a function of temperature and field has been studied in detail by point contact spectroscopy (PCS) [7-9] and by scanning tunneling spectroscopy (STS) [10-12].

The critical temperature and other superconducting properties of a two-band superconductor depend on the doping level and on the interband and intraband scattering [13-15] and therefore, they can be modified by chemical substitutions. Substitutions change the electronic structure, superconducting gaps, the defect structure, the inter- and intraband scattering, and thus superconducting properties such as $T_c$, upper critical fields, $H_{c2}$, and their anisotropy. Of potential practical interest would be an enhancement of the upper critical field. However, modifications of the properties by chemical substitutions in $MgB_2$ are still not well understood.

In order to study the influence of intra- and interband scattering on the gap and other superconducting properties investigations of partially substituted $MgB_2$ crystals are of particular interest. So far substitutions in $MgB_2$ were studied mainly on polycrystalline samples or thin films. Due to the anisotropic character of $MgB_2$, single crystal studies provide detailed insight and such studies have been reported for C, Al and Mn substitutions [16-21]. Other substitutions, such as Li [28,29] and co-substitutions with Al and Li [30,31] were investigated on polycrystalline samples or thin films.

Carbon substitution for boron produces particularly pronounced modifications: the upper critical field for both orientations $H_{c2}^{\|c}$ and $H_{c2}^{\|ab}$ increase with an associated decrease of the $H_{c2}$ anisotropy $\gamma = H_{c2}^{\|ab}/H_{c2}^{\|c}$ [18,22,23]. This can be explained by a mean free path reduction due to increased intraband scattering in the $\sigma$ band [18,24,25,26]. Carbon introduces defects in the $MgB_2$ structure causing a broadening of the x-ray reflections and a change of the flux pinning strength for low fields.

The length scale of such defects caused by inhomogeneous C distribution is likely shorter than the coherence length $\xi$, because otherwise magnetization curves should show multi-step or very broad transitions. Aluminum substitution for magnesium decreases $H_{c2}$ for both orientations of the field $\|c$ and $\|ab$, except for a small increase of $H_{c2}^{\|c}$ for low doping level, and decreases $H_{c2}$ anisotropy, making the anisotropy less temperature dependent. The C substitution for B as well as the Al substitution for Mg add electrons to $MgB_2$. Theory predicts merging of both gaps with increased interband scattering caused by the doping or impurities [5,15]. Such effect was confirmed experimentally for $MgB_2$ with 13% C substitution for B [27]. According to Kortus *et al.* [5], stronger interband scattering for C than that for Al substitutions can be explained by the fact that $\sigma$ band orbitals are located in the B plane, and there is not much weight of the $\sigma$ band in the Mg plane. The $\pi$ orbitals are also centered at the B plane, however extend further out towards the Mg plane. For these reasons impurities in the B plane are much more effective interband scatterers than impurities in the Mg plane. The decrease of $T_c$ with Al or C substitutions has been attributed to σ hole-band filling. Carbon and aluminum electron dopants reduce the number of holes at the top of the σ band together with the reduction of the electronic DOS and thus both decrease $T_c$ with a similar rate. Mn is found to substitute isovalently for Mg and the Mn magnetic moment interacts strongly with the conduction electrons and leads to an effective magnetic pair breaking. This results in a strong suppression of superconductivity ($T_c = 0$ for 2% of Mn) and its anisotropy [20].

Hole doping in MgB$_2$ has been investigated very little and only few papers concerned MgB$_2$:Li have been published [28,29]. To the best of our knowledge no single crystal results were reported so far for Li substitution. Co-doping with holes and electrons is a very interesting issue, because it can bring new information about the electronic band doping and intra- and interband scattering [5]. In the crystals where Li and C or Li and Al are substituted simultaneously one can expect compensation of the electron doping effect and an increase of $T_c$. Two papers have been published on the Al-Li co-substitution in polycrystalline samples [30,31], which show that there is no effect of Li content on $T_c$ and the value of $T_c$ depends only on Al content. This led to the conclusion, that in this case not only the band filling but also the lattice distortion plays an important role. On the other hand, the paper by Bernardini *et al.* [32] point out the significance of the effect of carrier doping. The authors calculated the effect of co-substitution Li-Al on the electronic structure and concluded that holes added by Li go almost entirely to the π band and thus do not counterbalance the electron donation from Al, which fill the σ band [32]. In such a case the observed changes in $T_c$ can be attributed to the effect of band filling alone.

In this paper we present the effect of the Li substitution for Mg and of the Li-C co-substitution on the superconducting properties and structure of MgB$_2$ single crystals. The objectives of these studies were twofold: First, we investigated the influence of Li$^{+1}$ substitution on $T_c$, structure, and $H_{c2}$ and its anisotropy. We show, that hole doping with Li decreases $T_c$, but in much slower rate than electron doping with C or Al. Second, we studied the role played by the Li doping in the Li-C co-substituted crystals and a possible counterbalance effect of simultaneous hole and electron doping. Such effect could have a practical importance because it might prevent $T_c$ to decrease without loosing the benefit of increased $H_{c2}$ due to the C substitution. We show that the effect of Li-C co-substitution on $T_c$ is different than that observed for Li-Al co-substitution [30-32], and Li added to C substituted MgB$_2$ decreases $T_c$ additionally.

## II. EXPERIMENT

Single crystals of Mg$_{1-x}$Li$_x$(B$_{1-y}$C$_y$)$_2$ were grown under high pressure using the cubic anvil press. The applied pressure/temperature conditions for the growth of MgB$_2$ single crystals were determined in our earlier study of Mg-B-N phase diagram [33,34]. Magnesium (Fluka, 99.99% purity), amorphous boron (Alfa Aesar, >99.99%), carbon graphite powder (Alfa Aesar, >99.99%) and lithium nitride (Alfa Aesar, >99.5%) were used as starting materials. Amorphous boron was annealed under dynamic vacuum at 1200 ºC to minimize contamination by oxygen. Due to extremely hydroscopic nature of lithium nitride, starting materials with various nominal contents were mixed and pressed in a glove box. A pellet was put into a BN container of 8 mm inner diameter and 8.5 mm length. The heating element was a graphite tube. Six anvils generate pressure on the whole assembly. Lithium and Li-C substituted crystals were grown in the same way as the unsubstituted crystals [33,34]. First, pressure of 30 kbar was applied using a pyrophylite pressure transmitting cube as a medium, then the temperature was increased during one hour up to the maximum of 1900-1950 °C, kept for 30 min, and decreased during 1-2 hours. We obtained Li substituted and double Li and C substituted $MgB_2$ single crystals with dimensions up to 1.5x0.8x0.1 $mm^3$ (Fig. 1). Crystals substituted with C were black in color in contrast to crystals substituted with Li or nonsubstituted, which were golden.

The carbon content in the crystals was estimated from the changes in the *a* lattice parameter. The lithium content was determined from structure refinement. Details of the structure investigations are described in the following chapter. In order to determine $T_c$, the magnetic moment of an individual crystal was measured at 0.1-0.5 mT field on a homemade SQUID magnetometer with a Quantum Design sensor. The magnetic measurements of the upper critical field were performed on a Quantum Design magnetic property measurement system (MPMS) with a 7 T magnet. Resistivity measurements were carried out in a Quantum Design physical property measurements system (PPMS) with a 14 T magnet using a standard four-point probe technique.

### III. RESULTS AND DISCUSSION

#### A. Single-crystal structure analysis

Several crystals with different Li content were investigated on single crystal x-ray CCD (in the Laboratory of Crystallography ETHZ) and Siemens P4 diffractometers. The unit cell parameters were estimated for each sample from the same set of 32 reflections in the wide 2Θ range (20-40 deg). Data reduction and analytical absorption correction were introduced using the CrysAlis software package [35].

The Li content was estimated using a refinement on $F^2$ [36]. Theoretical calculations indicate that only 0.75 of the magnesium atoms per formula unit are needed to provide the chemical bonding in $Mg_{0.96}B_2$ [37]. The rest of the atoms (0.96 – 0.75 = 0.21) gives additional electrons responsible for covalent interactions between magnesium and boron atoms along [001] direction [38]. It was assumed that lithium ions occupy the Mg site and provide fewer electrons than the Mg ions. Consequently, the sum of both cations was increased to 100%. The positions and atomic displacement parameters (ADP) for both cations were held to be equal (restrained). The refinement data, presented in Tabs. 1 and 2, confirm that Li enters Mg site. The estimated standard deviations of the lithium concentration were about 1-1.5%. Figure 2 shows the variation of the lattice parameters with Li content. The lattice parameter *c* decreases with Li substitution, while *a* appears to remain essentially constant, after an initial decrease at low Li concentrations. The carbon content in Li-C co-substituted crystals was estimated from the changes of the lattice parameter *a*, assuming the linear dependence of the *a* parameter on the carbon content, according to the data of Avdeev *et al.* [39], and by taking into account the decrease of the *a* due to the presence of Li. We assumed that C substitutes only B. The occupations of anions were held constant with the sum of both anions equal to 100%. The positions and atomic displacement parameters of both anions were held to be equal (restrained).

The two-dimensional profiles of the (002) reflection for the samples with a different composition ($MgB_2$, $Mg_{0.89}Li_{0.11}B_2$, $Mg_{0.89}Li_{0.11}B_{1.88}C_{0.12}$ and $Mg_{0.90}Li_{0.10}B_{1.85}C_{0.15}$) were constructed from the 30 - 40 one-dimensional scans using P4 Siemens diffractometer (Fig. 3). In pure as well as in Li substituted $MgB_2$ the reflection profiles are narrow in the direction perpendicular to the *c\** axis but elongated along *b\** in the *a\*b\** plane of the reciprocal space (Fig. 3 a and b). Such anisotropic broadening of reflections indicates disorder. For the Li substituted crystal this elongation is smaller (Fig. 3 b, bottom picture) than for the pure $MgB_2$ (Fig. 3 a, bottom picture). Unsubstituted $MgB_2$ crystals are deficient in Mg by about 4-5%. Most probably substitution of Li fills Mg vacancies and crystals are more ordered.

As grown samples substituted with the carbon and lithium show very broad (002) reflections (Fig. 3 c), most likely due to disorder. In comparison with the as grown carbon and lithium substituted crystals, samples with almost the same carbon and lithium content annealed in situ at 1800 °C for 2 h (after crystal growth) show much narrower reflections profiles (Fig. 3 d). Their shapes are similar to the reflections profiles of the pure $MgB_2$ and Li doped $MgB_2$ (Fig. 3 a, b). This observation agrees with superconducting transition width, $\Delta T_c$, dependence as a function of $T_c$ presented in the following chapter. The values of $\Delta T_c$ for pure $MgB_2$, for Li doped, and for annealed Li-C substituted crystals are much smaller than that one for as grown $Mg_{0.89}Li_{0.11}B_{1.88}C_{0.12}$, what indicates strong disorder in the last one.

The study of the physical properties of the $MgB_2$ crystals as a function of composition demands particular attention for the thermal treatment on the final stage of synthesis.

## B. Magnetic investigations

Figure 4 shows the normalized diamagnetic signal in the vicinity of $T_c$ for $Mg_{1-x}Li_xB_2$ crystals with various Li content. The superconducting transition temperature was determined from the magnetic moment measurements performed as a function of temperature in a 0.5 mT dc field in zero field cooling (ZFC) mode. The effective transition temperature, $T_c$, and the onset temperature, $T_{c,on}$, were defined as illustrated in Fig. 4. Figure 5 shows the $T_c$ dependence on the Li content, revealing that up to 8% of the Li content $T_c$ decreases very little, which can correspond to Mg vacancy filling. On the other hand, the observed effect can be explained assuming that as grown $MgB_2$ crystal is slightly underdoped and its $T_c$ is very insensitive for the doping with small amount of holes. Above 8% of the Li content a sharp drop of $T_c$ is observed, which may indicate appearance of new structural defects or strongly increasing contribution of existing defects to the intra- or interband scattering. The sharp drop of $T_c$ can be explained also as an effect of reaching the concentration of carrier significantly higher than that one corresponding to maximum of $T_c$.

The $T_c$ dependence as a function of content of various substitutions in $MgB_2$ for our single crystals are compared in Fig. 6 [18-20]. The rate of the $T_c$ reduction is similar for C and Al substitutions. For the Mn substitution the change is much more rapid because of magnetic pair breaking. For the Li substitution, $T_c$ also decreases with the increasing Li content, however the rate is much slower than that one ~~in~~ for C or Al substituted crystals. The strongly non-symmetric change of $T_c$ due to the substitution with Li and with C may indicate that as grown $MgB_2$ crystals are slightly underdoped. Doping with holes introduced by the Li substitution may eventually lead to a slight increase of $T_c$, however this effect is overcompensated by an introduction of a lattice distortion acting as scattering centers in the substituted crystals.

In the $MgB_2$ crystals co-substituted simultaneously with both Li and C one can expect that the hole doping with Li will compensate the effect of the electron doping with C. The $T_c$ value for $Mg_{1-x}Li_x(B_{1-y}C_y)_2$ may increase in comparison with $T_c$ for $Mg(B_{1-y}C_y)_2$. In order to verify the above prediction, three sets of substituted $MgB_2$ crystals were investigated. For two of them, the crystals were substituted separately with one element only, i.e., with C for B or with Li for Mg, and for one set, the crystals were substituted with both C and Li. The crystals substituted with both C and Li have lower $T_c$ than the crystals substituted with the same amount of C only (Fig.7). For example, if we compare the crystals with 6% of C and with 6% of C + 11% of Li, we find a difference in $T_c$ of about 3.3 K. Almost the same difference in $T_c$ is seen in Fig. 5 comparing $T_c$ data for 12% of Li substituted and for unsubstituted crystals.

In order to draw more extended picture of the $T_c$ dependence on the C and Li content we have investigated crystals with a various substitution level. Figure 8 shows a summary of our results for the C substitution from 2% up to 9% and the simultaneous Li substitution from 4% up to 12%. In the whole range of substitutions, $T_c$ is lower for C and Li co-substituted than for C only substituted crystals. Dashed lines show the results for crystals with the same level of the Li substitution for various C content. The same amount of Li substitution in $Mg(B_{1-x}C_x)_2$ crystals leads to very similar decrease of their $T_c$, even for the crystals with quite different carbon content.

All the discussed results indicate that changes in the carrier concentration caused by holes introduced with Li do not counterbalance the decrease of $T_c$ due to C substitution. It can be explained assuming that holes introduced by Li occupy almost exclusively the $\pi$ band and do not fill the $\sigma$ band, thus cannot compensate the electrons donated from C which fill the $\sigma$ band [32]. In other words, substituted Li does not change the filling of the $\sigma$ bands but accepts electrons from the $\pi$ band. It is known, that the $\sigma$ band is responsible for superconductivity in $MgB_2$ and filling of the $\pi$ band should not affect $T_c$. The question appears, however, what the reason of $T_c$ decreasing by the Li substitution is? If the picture of isolated impurities is right the observed decrease can be caused most likely by enhanced scattering due to defects introduced by substitution and due to distortion of the lattice. In order to verify this prediction some $(Mg,Li)(B,C)_2$ crystals have been annealed after the growth process at high pressure at 1800°C for 1.5-2 hours. We expected that in this way the lattice distortion may be reduced. In fact, as one can see in Fig. 8, for the part of the annealed crystals with C content $x$ = 0.03-0.04 and with Li content of 0.10-0.12, the values of $T_c$ are similar to those obtained for the $MgB_2$ crystals substituted with C only. However, for higher C content, $x$ = 0.06-0.08, this effect is not so obvious, which indicates that for heavier C substitution, the annealing does not remove the lattice distortion. Electronic structure investigations by point contact spectroscopy or by STM may be useful to explain this behavior.

The effect of substitutions on disorder can be also observed in Fig. 9, where $\Delta T_c$ as a function of $T_c$ of substituted crystals is shown. Substitution with Li and C and co-substitution with Li-C increases $\Delta T_c$ significantly. Annealing of Li-C co-substituted crystals decreases $\Delta T_c$ what is in agreement with the decrease of reflection width as a result of annealing shown in Fig. 3.

The upper critical field $H_{c2}$ has been determined from magnetic moment measurements performed as a function of temperature at constant magnetic field or at various magnetic fields at constant temperature. Figure 10 shows examples of $M(T)$ dependences for various magnetic fields. The results have been obtained with a field oriented parallel to the *ab* plane (not shown in the Figure), $H \parallel ab$, and parallel to the *c* axis, $H \parallel c$. The difference between $T_c$ and $T_{c,on}$ increases slightly with increasing field, but usually do not exceed 1 K. Sets of the data similar to these presented in Fig. 10 were used to construct the $H_{c2}$-$T$ phase diagram for the samples with various Li and Li-C content. Figure 11 shows the phase diagram for two $MgB_2$ crystals substituted with Li and, for comparison, for an unsubstituted crystal. These results clearly show that $H_{c2}^{\parallel ab}$ decreases with Li doping, while $H_{c2}^{\parallel c}$ does not change significantly. Therefore, the resulting upper critical field anisotropy, $\gamma$, decreases with increasing Li content. An increase of the intraband scattering with Li doping may play a role as a source of observed changes [13], because for *H* parallel to the *ab*-plane, a substantial decrease of $\mu_0 dH_{c2}/dT$ at $T_c$ from 0.20 T/K for unsubstituted MgB$_2$ to 0.16 T/K for Mg$_{1-x}$Li$_x$B$_2$ with $x = 0.11$ was observed.

Temperature dependence of $H_{c2}$ for Al and Li substituted crystals is shown in Fig. 12. The crystals with similar $T_c$ show an almost identical $dH_{c2}/dT$ slope at $T_c$ and a similar $H_{c2}(T)$ dependence, despite of much different substitution levels for different substitutes. It indicates that the mechanism controlling $H_{c2}$ is very similar to the mechanism determining $T_c$ for both Li and Al substituted crystals. Two effects can be responsible for the changes in $T_c$ and $H_{c2}$. The first one is the doping effect due to the changes in the carrier concentration, and the second one is the introduction of new scattering centers, enhancing the interband and/or intraband scattering and changing the superconducting gaps.

The upper critical field $H_{c2}$ for the double-substituted Mg$_{1-x}$Li$_x$(B$_{1-y}$C$_y$)$_2$, and single-substituted Mg$_{1-x}$Li$_x$B$_2$, and Mg(B$_{1-y}$C$_y$)$_2$ crystals is shown in Fig. 13. For $H$ parallel to the $ab$-plane, significantly different amount of substituted C results in a similar $T_c$ and similar $H_{c2}(T)$ dependence, slightly shifted into higher temperatures for the $x = 0.06$ and $y = 0.02$ crystal as compared with that one for the $x = 0.0$ and $y = 0.05$ crystal. This is a little surprising result, because for MgB$_2$ single-substituted with Li, both $T_c$ and $H_{c2}(T)$ are only somewhat influenced by introduced Li up to the content of $x = 0.08$ (see Fig. 11). Considerable changes of the superconducting properties of Mg$_{1-x}$Li$_x$B$_2$ crystals are observed for the composition with $x = 0.11$. However, for the double-substituted crystals, Mg$_{1-x}$Li$_x$(B$_{1-y}$C$_y$)$_2$, the influence of Li is much more significant. Crystals with $x = 0.06$ and $y = 0.02$ has $T_c = 35.2$ K, while $T_c = 37$ K is expected for Mg(B$_{1-y}$C$_y$)$_2$ with $y = 0.02$. So substantial change of $T_c$ from 37 to 35 K due to the substitution of only 6% of Li is a real puzzle. We can speculate, that this is effect of increasing role of defects acting as inter- and/or intraband scattering centers when specific critical concentration of the defects in the superconductor is reached. However, the results show unambiguously that the effect of co-substitution on the electronic structure and superconducting properties of MgB$_2$ cannot be treated as a trivial one and needs to be studied deeply.

## C. Electrical transport measurements

Figure 14 shows the normalized $ab$-plane resistance $R(T)/R(300)$ measured on MgB$_2$ single crystals substituted with Li and co-substituted with Li and C. One of the co-substituted crystals has been post annealed at high pressure at 1800 °C, after completing the growing process. The data obtained for MgB$_2$ and for C-substituted MgB$_2$ are presented for comparison as well. It is clearly visible that the normalized resistance increases with increasing doping level and that the introduction of Li into the structure affects the in-plane transport properties by increasing the scattering of charge carriers. This effect is noticeable in the case of the Li substituted crystal as well as in the case of the Li-C co-substituted ones.

A similar trend can be observed in Fig. 15, where the absolute values of the resistivity are shown. The residual resistivity $\rho_0$ increases from ~ 1.2 $\mu\Omega$ cm for unsubstituted MgB$_2$ up to ~ 5.6 $\mu\Omega$ cm for the Li substituted crystal, indicating an increasing amount of defects and thus larger disorder. For Mg$_{1-x}$Li$_x$(B$_{1-y}$C$_y$)$_2$ with $x=0.12$ and $y=0.03$, $\rho_0$ is ~ 8.3 $\mu\Omega$ cm. A similar value of $\rho_0$ (~ 8.9 $\mu\Omega$ cm) has been obtained for the crystal with $x=0.06$ and $y=0.02$. For Mg$_{1-x}$Li$_x$(B$_{1-y}$C$_y$)$_2$ with $x=0.09$ and $y=0.08$, $\rho_0$ increases up to ~ 22 $\mu\Omega$ cm. Resistivity measurements for C substituted crystal with $y=0.05$ ($\rho_0$ ~ 9.9 $\mu\Omega$ cm) and for the crystal with $y=0.083$ ($\rho_0$ ~ 13 $\mu\Omega$ cm) are presented in the figure for comparison. The uncertainty in the estimation of the resistivity due to the geometrical factor was ~ 15%.

A comparison between the crystal substituted with 8.3% of C and the co-substituted one with 8% of C and 9% of Li shows that, for a practically equal amount of C, $\rho_0$ is sensitively higher in the co-substituted crystal than in the crystal substituted with C only. This means that, for the crystals substituted with C, the additional substitution with Li increases the amount of defects resulting in a higher $\rho_0$.

Furthermore, comparison of the resistivity for the crystal with $x=0.06$ and $y=0.02$ and for the annealed one with $x=0.12$ and $y=0.03$, indicate that, for the crystals with a similar amount of C (2% and 3%) but with a significantly different Li content (6% and 12%), the in-plane resistivities are almost equal if the crystal with higher Li content is relaxed. This observation indicates that, for particular level of C substitution the disorder introduced by substitution of Li can be removed, at least partially, by annealing. The relaxation process manifests itself in the fact that $T_c$ for the annealed crystals is higher than that one for the non-annealed ones.

The upper critical field has been evaluated from the electrical resistance measurements, for $H$ parallel to both the $ab$ plane and the $c$ axis. The in-plane resistivity $\rho_{ab}$ of the crystal substituted with 2% of C and 6% of Li is presented in the inset of Fig. 14 for several values of $H$ oriented parallel to the $c$ axis. The sharp resistive transition observed in $H = 0$ gradually broadens with increasing field showing a two-step-like characteristics. This behavior, significantly enhanced in the field applied along the $c$ axis, can be attributed to vortex melting and has been observed also for unsubstituted [40] and C substituted $MgB_2$ [18]. In order to estimate the upper critical field we adopted two different definitions of $T_c$, temperature of the onset of resistivity drop and temperature of the appearance of zero resistivity (see inset of Fig. 14). In principle this two temperatures should define respectively the upper critical field and the irreversibility field [41,42]. In the case of the field parallel to the $c$ axis these two temperatures may differ very much since $T_{c,on}$ can be affected by surface effects [40]. Therefore, it seems to be more reasonable to define $H_{c2}$ by means of zero resistivity criterion, leading to $T_c$ in agreement with several bulk measurements performed on pure $MgB_2$ [6,40,43,44].

The upper critical fields, $H_{c2}^{\|ab}$ and $H_{c2}^{\|c}$, obtained with zero resistivity criterion from the in-field resistivity measurements (see the inset of Fig. 14) performed for Li substituted and for Li-C co-substituted crystals are shown in Fig. 16. In this Figure, $H_{c2}^{\|c}$ evaluated with $T_{c,on}$ is also presented and, as expected, it is more temperature dependent than $H_{c2}^{\|c}$ determined by means of zero resistivity criterion. The upper critical field anisotropy $\gamma$ is smaller for the crystal substituted with 12% Li and 3% C, due to the notable decrease of $H_{c2}^{\|ab}$ and practically unchanging $H_{c2}^{\|c}$.

The temperature dependence of the upper critical field anisotropy $\gamma$ for $MgB_2$ crystals substituted with Li or C and co-substituted with Li and C simultaneously is shown in Fig.17. The anisotropy decreases with increasing temperature for all compositions but is much lower than that one for unsubstituted $MgB_2$. A negative anisotropy slope, $d\gamma/dT$, is expected for the case when diffusivity in the $\pi$ band dominates [13]. This requirement seems to be fulfilled in the clean unsubstituted or C-substituted $MgB_2$, where C substitution on the B site leads to decrease of the diffusivity mainly in the $\sigma$ band, as shown for C substituted single crystals [25] and for epitaxial thin films [45]. On the other hand, when the diffusivity in the $\sigma$ band dominates, $\gamma(T)$ is expected to be less temperature dependent, or the slope of $d\gamma/dT$ may even become positive [13,46]. The diffusivity in the $\sigma$ band may dominate, when the scattering in the $\pi$ band increases substantially, e.g., due to the substitution of Li for Mg. The results shown in Fig. 17 indicate that the ratio of the intraband scattering rate in the $\pi$ band to the intraband scattering rate in the $\sigma$ band does not increase significantly for both the C substituted and Li-C co-substituted crystals. However, for the crystals with 11% of Li, some flattening of the temperature dependence of $\gamma$ is observed, similarly to that what was reported for Al substituted $MgB_2$ [19]. This means that while scattering in the $\sigma$ band may still dominate in the crystals co-substituted with Li and C, in the crystals substituted with Li the $\pi$ band scattering may contribute notable to the net scattering and thus influence the transport properties. The details can be worked out through $T$-dependent gap spectroscopy studies on substituted crystals.

## IV. Conclusions

The substitution of $Li^{+1}$ for $Mg^{+2}$ introduces holes and leads to the decrease of $T_c$ at a much lower rate (~ 0.3 K/%Li) than the substitution of $Al^{+3}$ for $Mg^{+2}$, which dopes electrons (~ 1.3 K/%Al). Co-substitution with Li and C decreases $T_c$ of $MgB_2$ crystals more than in the case where C is substituted alone. The possible reason of this feature can be that holes introduced with Li occupy the $\pi$ band and do not compensate the electrons introduced with C, which fill the $\sigma$ band. In that case holes doped by the substitution of Li should not influence $T_c$. The decrease of $T_c$ can be caused by an increase of the interband scattering. In Li and C co-substituted crystals $T_c$ decreases as a result of both electron doping and impurity scattering. An alternative explanation of the phenomena observed for $MgB_2$ crystals with various substitutions can be as follows: The reduction of $T_c$ due to the increased or decreased number of charge carriers seems to reveal that $MgB_2$ is close to an optimally doped compound. The strongly non-symmetric decrease of $T_c$ due to the substitution of Li and C may indicates, that in the phase diagram "carrier concentration-transition temperature" $MgB_2$ is located close to the maximum of $T_c$, in slightly underdoped region. Doping with electrons, introduced by C or Al substitutions, decreases $T_c$ as a result of a decrease of the hole-carrier content as well as a result of the introduction of new scattering centers. Doping with holes introduced by the Li substitution may eventually lead to a slight increase of $T_c$, however this effect is overcompensated by introduction of distortion acting as new scattering centers. Obtained data rule out utilization of the counterbalance effect of simultaneous hole and electron doping, which was expected to prevent $T_c$ from decreasing in the C substituted $MgB_2$ without loosing the benefit of increased $H_{c2}$.

X-ray investigations show the broadening of the reflections while magnetic investigations show increasing of $\Delta T_c$ in substituted crystals, which indicate structural disorder. In the crystals co-substituted with Li and C an increase of $T_c$, narrowing of x-ray reflections and decrease of $\Delta T_c$ is observed as a result of annealing at 1800 ºC at high pressure. Distortion of the lattice due to substitution appears to be a significant factor in the modification of $T_c$.

**Acknowledgments**

The authors are grateful to A. Mironov for helpful discussions. This work was supported by the Swiss National Science Foundation through NCCR pool MaNEP, the Polish State Committee for Scientific Research under a research project for the years 2004–2007 (1 P03B 037 27), and the Polish Ministry of Science and Higher Education under a research project for the years 2006-2009 (No. N202 131 31/2223).

**TABLE I.** Structure refinement and crystal data for $MgB_2$ and for $MgB_2$ doped with Li.

| Sample | AN307/17 | AN407/2 | AN453/5 | AN453/7 |
|---|---|---|---|---|
| Empirical formula | $Mg_{0.94}B_2$ | $Mg_{0.95}Li_{0.05}B_2$ | $Mg_{0.90}Li_{0.10}B_2$ | $Mg_{0.89}Li_{0.11}B_2$ |
| $T_{c,on}$, K | 38.8 | 38.5 | 35.6 | 35.8 |
| $T_c$, K | 38.75 | 38.25 | 35.55 | 35.3 |
| Formula weight | 44.47 | 45.06 | 44.19 | 44.02 |
| Temperature, K | 295(2) | | | |
| Wavelength, Å/radiation | 0.71073/$Mo$ K$\alpha$ | | | |
| Cell determined on | Siemens P4 four circles diffractometer (Point detector) | | | |
| 2$\Theta$ range for cell determination, deg | 31,1 | | | |
| Intensity collection on diffractometer | Oxford diffraction four circles diffractometer (CCD detector) | | | |
| Crystal system, space group | Hexahonal, $P6/mmm$ | | | |
| Unit cell dimensions, Å | $a$ = 3.0865(2), $c$ = 3.5208(7) | $a$ = 3.0837(4), $c$ = 3.5176(6) | $a$ =3.0828(8), $c$ =3.512(1) | $a$ = 3.0826(5), $c$ = 3.5109(8) |
| Unit cell volume, Å$^3$ | 29.047(9) | 28.968(7) | 28.91(2) | 28.89(1) |
| Z | 1 | | | |
| Calculated density, g/cm$^3$ | 2.543 | 2.584 | 2.539 | 2.531 |
| Absorption correction type | analytical | | | |
| Absorption coefficient, mm$^{-1}$ | 0.579 | 0.592 | 0.560 | 0.555 |
| $F$(000) | 21 | 22 | 21 | 21 |
| Crystal size, mm | 0.28 x 0.19 x 0.11 | 0.26 x 0.11 x 0.03 | 0.27 x 0.11 x 0.03 | 0.16 x 0.11 x 0.005 |
| $\Theta$ range for data collection, deg | 5.80 to 37.33 | 5.80 to 30.34 | 5.81 to 36.08 | 5.81 to 30.11 |
| Limiting indices | -5≤$h$≤4, -4≤$k$≤5, -4≤$l$≤6 | -4≤$h$≤4, -4≤$k$≤4, -4≤$l$≤5 | -5≤$h$≤5, -5≤$k$≤5, -5≤$l$≤5 | -4≤$h$≤4, -4≤$k$≤3, -4≤$l$≤3 |
| Reflections collected/unique | 221/47, Rint = 0.0187 | 477/33, Rint = 0.0423 | 1017/46, Rint = 0.0336 | 269/31, Rint = 0.0212 |
| Max. and min. transmission | 0.950 and 0.920 | 0.956 and 0.834 | 0.974 and 0.870 | 0.989 and 0.902 |
| Refinement method | Full-matrix least-squares on $F^2$ | | | |
| Data /restraints/parameters | 47/0/7 | 33/0/6 | 46/0/6 | 31/0/6 |
| Goodness-of-fit on $F^2$ | 1.176 | 1.202 | 1.173 | 1.286 |
| Final R indices [$I$>2sigma($I$)] | $R_1$ = 0.0261, w$R_2$ = 0.0650 | $R_1$ = 0.0306, w$R_2$ = 0.0778 | $R_1$ = 0.0233, w$R_2$ = 0.0585 | $R_1$ = 0.0183, w$R_2$ = 0.0540 |
| R indices (all data) | $R_1$ = 0.0271, w$R_2$ = 0.0651 | $R_1$ = 0.0309, w$R_2$ = 0.0778 | $R_1$ = 0.0278, w$R_2$ = 0.0590 | $R_1$ = 0.0196, w$R_2$ = 0.0542 |
| $\Delta\rho_{max}$ and $\Delta\rho_{min}$, (e/Å$^3$) | 0.145 and -0.425 | 0.216 and -0.461 | 0.299 and -0.289 | -0.222 and 0.203 |

**TABLE II.** Structure refinement and crystal data for $MgB_2$ substituted with Li and C.

| Sample | AN467/1 | AN456/6 | AN456/10 | AN456/4 |
|---|---|---|---|---|
| Empirical formula | $Mg_{0.94}Li_{0.06}B_{1.96}C_{0.04}$ | $Mg_{0.91}Li_{0.09}B_{1.84}C_{0.16}$ | $Mg_{0.90}Li_{0.10}B_{1.82}C_{0.18}$ | $Mg_{0.89}Li_{0.11}B_{1.88}C_{0.12}$ |
| $T_{c,on}$, K | 35.8 | 30.6 | 28.9 | 30.2 |
| $T_c$, K | 35.6 | 30.15 | 28.5 | 29.9 |
| Formula weight | 44.94 | 44.56 | 44.41 | 44.16 |
| Temperature, K | 295(2) | | | |
| Wavelength, Å/radiation | 0.71073/*Mo* Kα | | | |
| Cell determined on | Siemens P4 four circles diffractometer (Point detector) | | | |
| 2Θ range for cell determination, deg | 31,1 | | | |
| Intensity collection on diffractometer | Oxford diffraction four circles diffractometer (CCD detector) | | | |
| Crystal system, space group | Hexahonal, *P6/mmm* | | | |
| Unit cell dimensions, Å | $a$ =3.075(2), $c$ =3.522(3) | $a$ =3.0561(6), $c$ =3.5190(10) | $a$ =3.053(1), $c$ =3.522(2) | $a$ =3.0602(8), $c$ =3.5243(9) |
| Unit cell volume, Å$^3$ | 28.84(4) | 28.46(1) | 28.43(2) | 28.58(2) |
| Z | 1 | | | |
| Calculated density, g/cm$^3$ | 2.588 | 2.601 | 2.595 | 2.567 |
| Absorption correction type | analytical | | | |
| Absorption coefficient, mm$^{-1}$ | 0.585 | 0.577 | 0.573 | 0.563 |
| F(000) | 22 | 21 | 21 | 21 |
| Crystal size, mm | 0.45 x 0.20 x 0.05 | 0.34 x 0.21 x 0.05 | 0.56 x 0.25 x 0.03 | 0.47 x 0.27 x 0.04 |
| Θ range for data collection, deg | 5.79 to 27.72 | 5.80 to 35.02 | 5.79 to 37.26 | 5.79 to 36.38 |
| Limiting indices | -3≤$h$≤3, -3≤$k$≤3, -4≤$l$≤4 | -4≤$h$≤4, -4≤$k$≤4, -5≤$l$≤5 | -5≤$h$≤5, -5≤$k$≤5, -5≤$l$≤6 | -5≤$h$≤5, -5≤$k$≤5, -5≤$l$≤5 |
| Reflections collected/unique | 407/25, Rint = 0.0543 | 478/44, Rint = 0.0268 | 1241/47, Rint= 0.0367 | 509/46, Rint = 0.0285 |
| Max. and min. transmission | 0.934 and 0.707 | 0.942 and 0.759 | 0.963 and 0.729 | 0.949 and 0.681 |
| Refinement method | Full-matrix least-squares on $F^2$ | | | |
| Data /restraints/parameters | 25/0/6 | 44/0/6 | 47/0/6 | 46/0/6 |
| Goodness-of-fit on $F^2$ | 1.300 | 1.259 | 1.285 | 1.290 |
| Final R indices [$I$>2sigma($I$)] | $R_1$ = 0.0245, $wR_2$ = 0.0700 | $R_1$ = 0.0343, $wR_2$ = 0.0895 | $R_1$ = 0.0254, $wR_2$ = 0.0649 | $R_1$ = 0.0376, $wR_2$ = 0.1015 |
| R indices (all data) | $R_1$ = 0.0245, $wR_2$ = 0.0700 | $R_1$ = 0.0356, $wR_2$ = 0.0906 | $R_1$ = 0.0274, $wR_2$ = 0.0654 | $R_1$ = 0.0395, $wR_2$ = 0.1023 |
| Δρ$_{max}$ and Δρ$_{min}$, (e/Å$^3$) | 0.160 and -0.289 | 0.292 and -0.438 | 0.276 and -0.234 | 0.368 and -0.426 |

Figures:

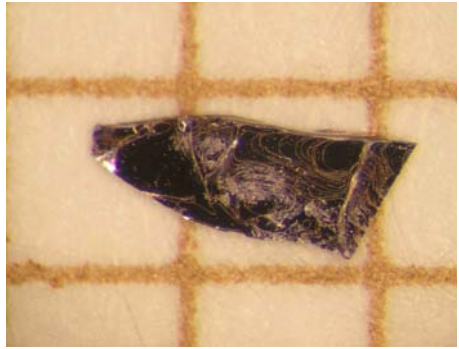

Fig. 1. (Color online) MgB$_2$ crystal substituted with Li. Scale is 1 mm.

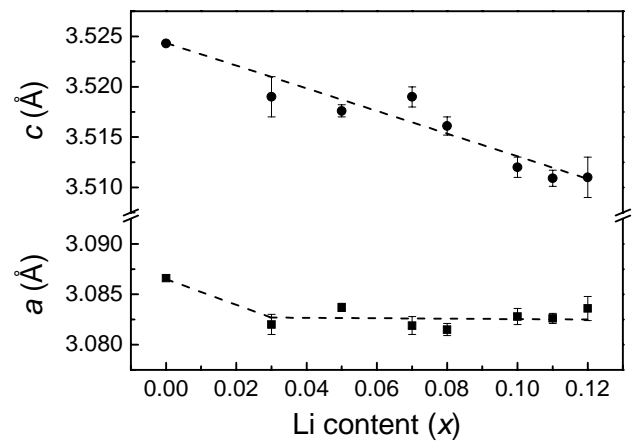

Fig. 2. Lattice parameters $a$ and $c$ as a function of Li content in Mg$_{1-x}$Li$_x$B$_2$ crystals.

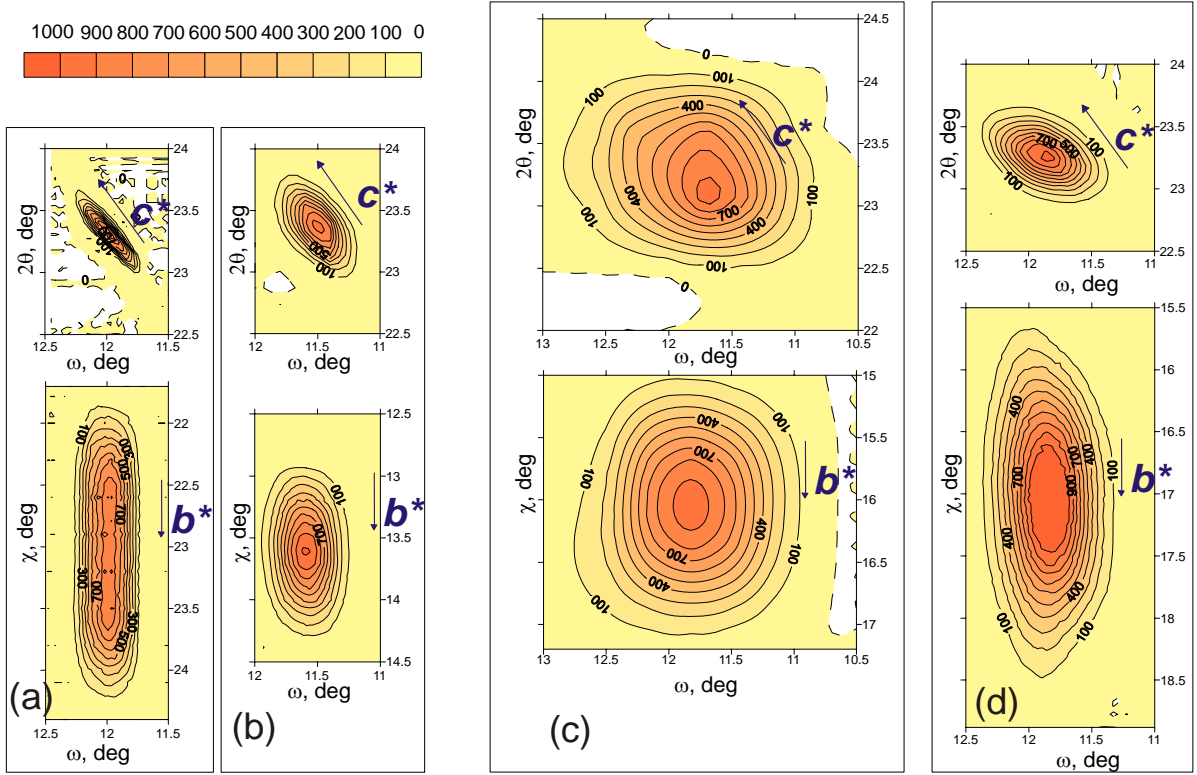

Fig. 3. (a), (b), (c) and (d) for MgB$_2$, Mg$_{0.89}$Li$_{0.11}$B$_2$, Mg$_{0.89}$Li$_{0.11}$B$_{1.88}$C$_{0.12}$ and Mg$_{0.90}$Li$_{0.10}$B$_{1.85}$C$_{0.15}$ annealed at 1800 °C for 2 h after crystal growth, respectively. Upper pictures: $\omega - \theta$ scan of the (002) reflections; $c^*$ is parallel to the reflection, and $a^*b^*$ is perpendicular to the plane. Lower pictures: $\omega - \chi$ scan of the (002) reflection; $c^*$ is perpendicular to the plane, $a^*b^*$ is parallel to the plane, and $b^*$ is parallel to the reflection. The scale is the same in all figures.

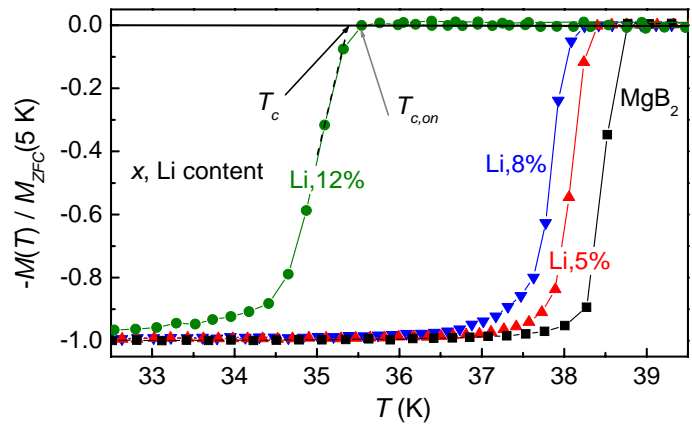

Fig. 4. (Color online) Normalized diamagnetic moment $M$ vs. temperature for the Mg$_{1-x}$Li$_x$B$_2$ single crystals with various Li content $x$. The measurements were performed in a field of 0.5 mT, after cooling in a zero field. The superconducting transition, $T_c$, and temperature of the transition onset, $T_{c,on}$, are marked with arrows.

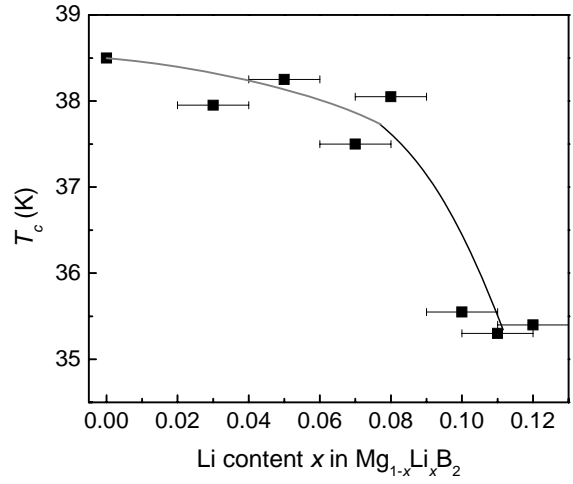

Fig. 5. Superconducting transition temperature $T_c$ dependence on the Li content in Mg$_{1-x}$Li$_x$B$_2$ single crystals.

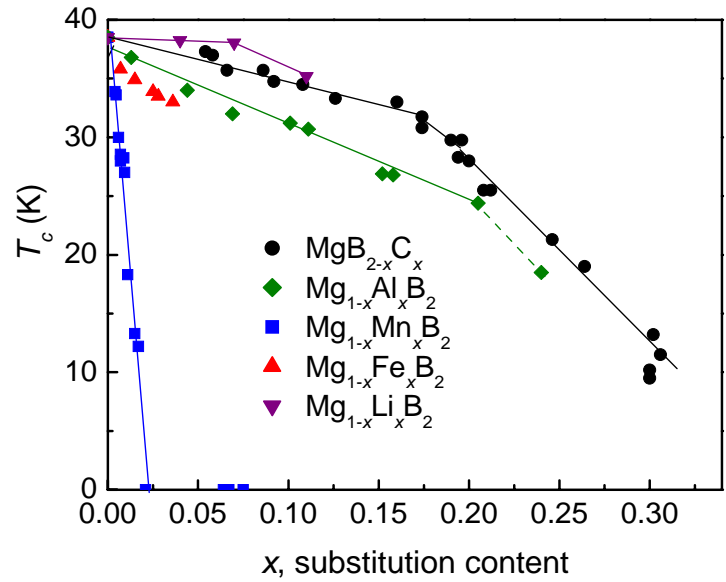

Fig. 6. (Color online) $T_c$ as a function of content of various substituents in MgB$_2$ [18-20]. Aluminum and carbon dope MgB$_2$ with electrons. Li$^{+1}$ dopes MgB$_2$ with holes, while isovalent Mn$^{+2}$ is a magnetic ion.

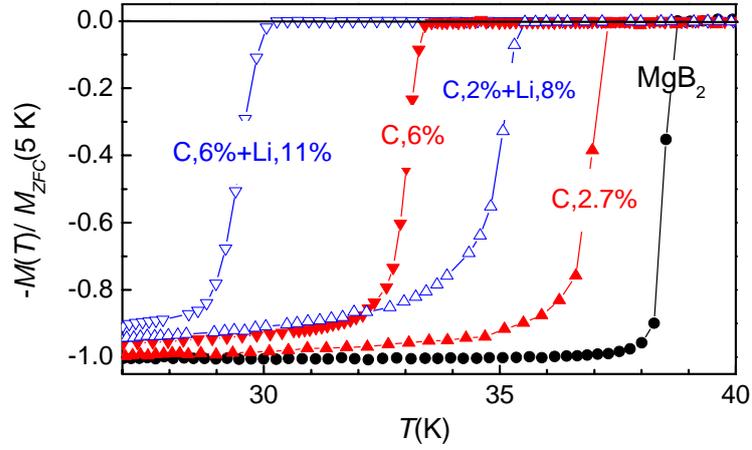

Fig. 7. (Color online) Normalized magnetic moment *M* vs. temperature for the crystals of $MgB_2$, substituted with C, and co-substituted with C and Li.

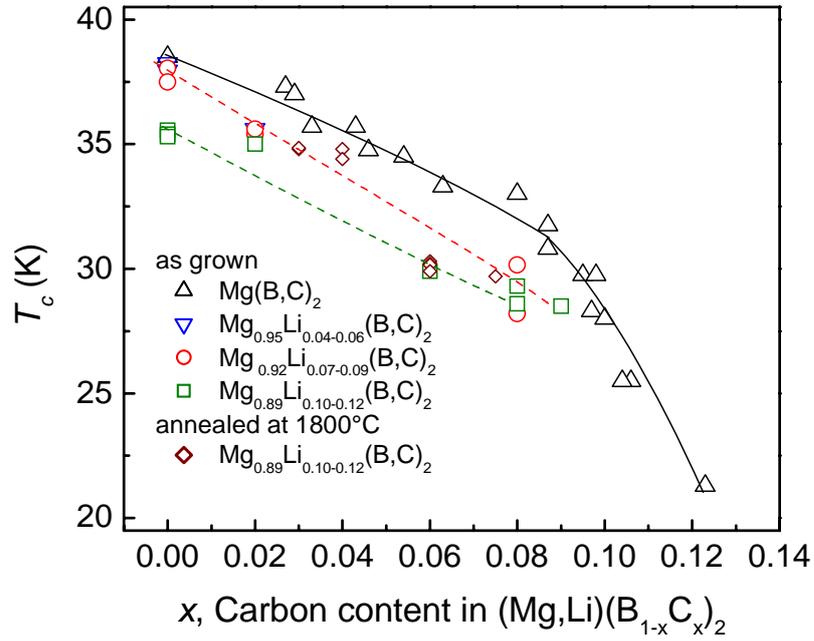

Fig. 8. (Color online) $T_c$ dependence on substitution of C (triangles) or co-substitution of C and Li (reversed triangles, circles, and squares). Dash lines show the same level of Li content for various level of C substitution.

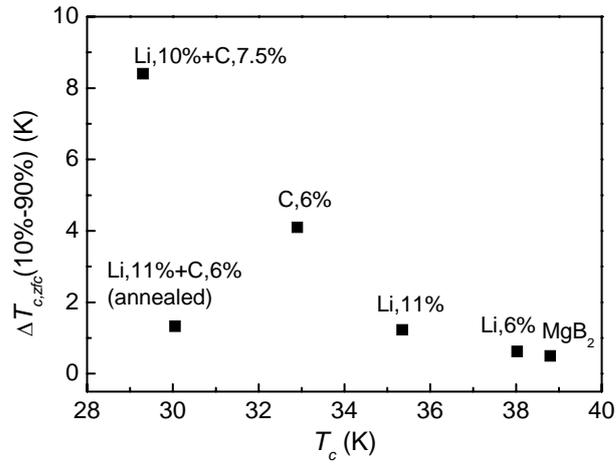

Fig. 9. Broadening of the superconducting transition $\Delta T_c$ for the pure $MgB_2$, for the crystals substituted with Li and C, for the crystals co-substituted with Li-C and for annealed Li-C co-substituted crystals of $MgB_2$ as a function of $T_c$.

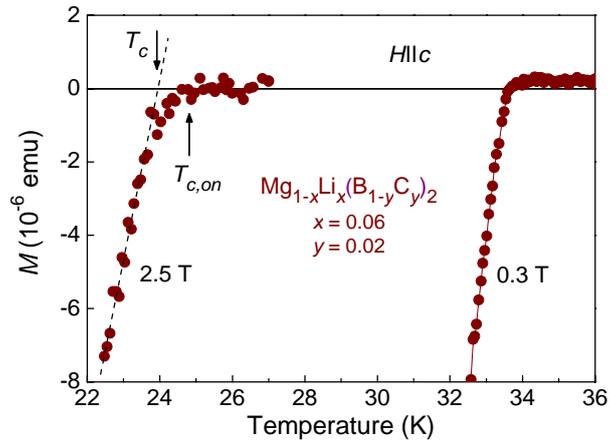

Fig. 10. Temperature dependence of magnetic moment in $H$ = 0.3 and 2.5 T parallel to the $c$-axis of the $Mg_{0.94}Li_{0.06}(B_{0.98}C_{0.02})_2$ crystal in the vicinity of $T_c$. $T_c$ and $T_{c,on}$ correspond to the transition temperature and to the temperature of the transition onset, respectively.

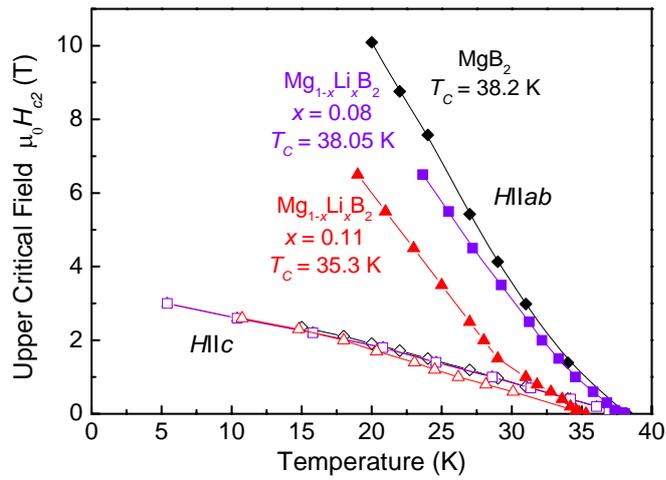

Fig. 11. (Color online) Upper critical field for two MgB$_2$ crystals substituted with 8% and 11% Li, compared with the upper critical field for an unsubstituted crystal.

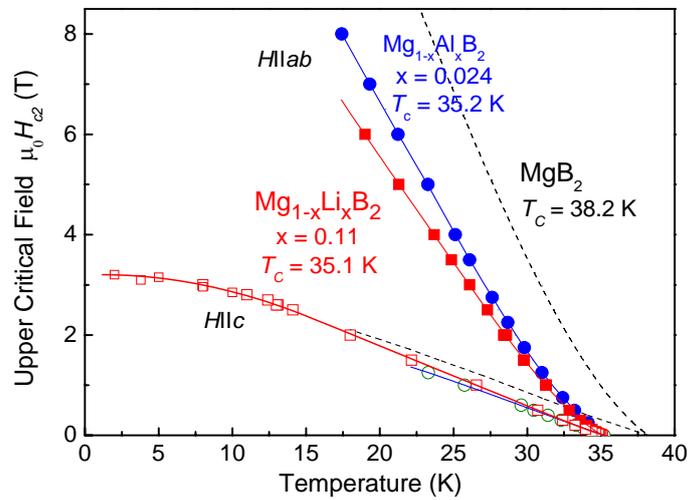

Fig. 12. (Color online) The upper critical field for Al and Li substituted crystals. Crystals with similar $T_c$ (but different substitution level for different substitutes) show a similar d$H_{c2}$/d$T$ slope at $T_c$ and a similar $H_{c2}(T)$ dependence.

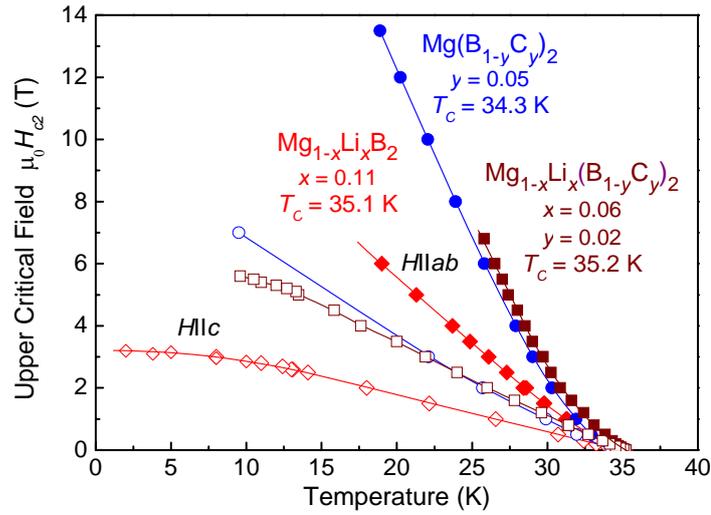

Fig. 13. (Color online) Upper critical field versus temperature for the double-substituted $Mg_{1-x}Li_x(B_{1-y}C_y)_2$ and single-substituted $Mg_{1-x}Li_xB_2$ crystals.

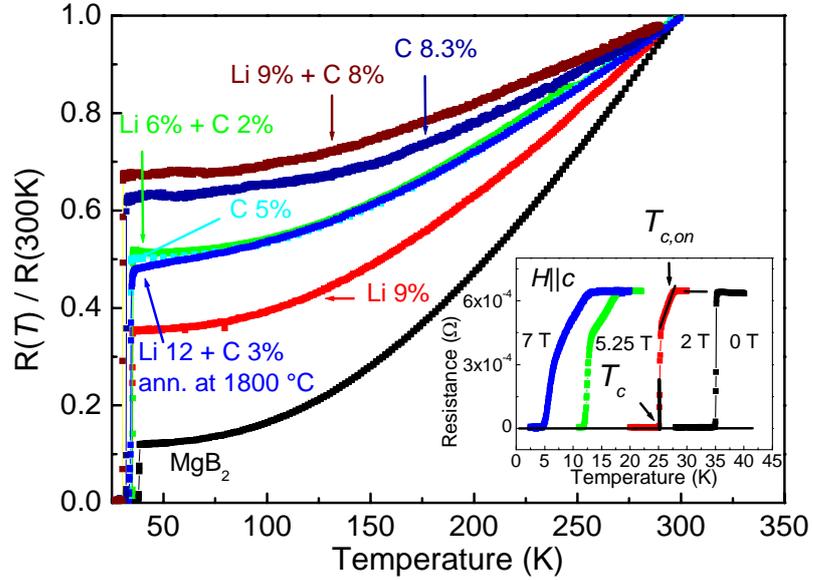

Fig. 14. (Color online) Normalized in-plane resistance $R(T)/R(300)$ as a function of temperature for single crystals of $Mg_{1-x}Li_xB_2$ with $x=0.09$ and of $Mg_{1-x}Li_x(B_{1-y}C_y)_2$ with $x=0.06$, $y=0.02$; $x=0.12$, $y=0.03$; and $x=0.09$, $y=0.08$. The crystal with $x=0.12$ and $y=0.03$ has been post annealed at high pressure at 1800 °C, after completing the growth. Data for $MgB_2$ and for $Mg(B_{1-y}C_y)_2$ with $y=0.05$ and $y=0.083$ are presented for comparison. Inset: resistive transitions measured for a co-doped single crystal with 6% of Li and 2% of C for various magnetic fields oriented parallel to the $c$ axis. The critical temperatures, $T_c$ and $T_{c,on}$, are indicated by arrows.

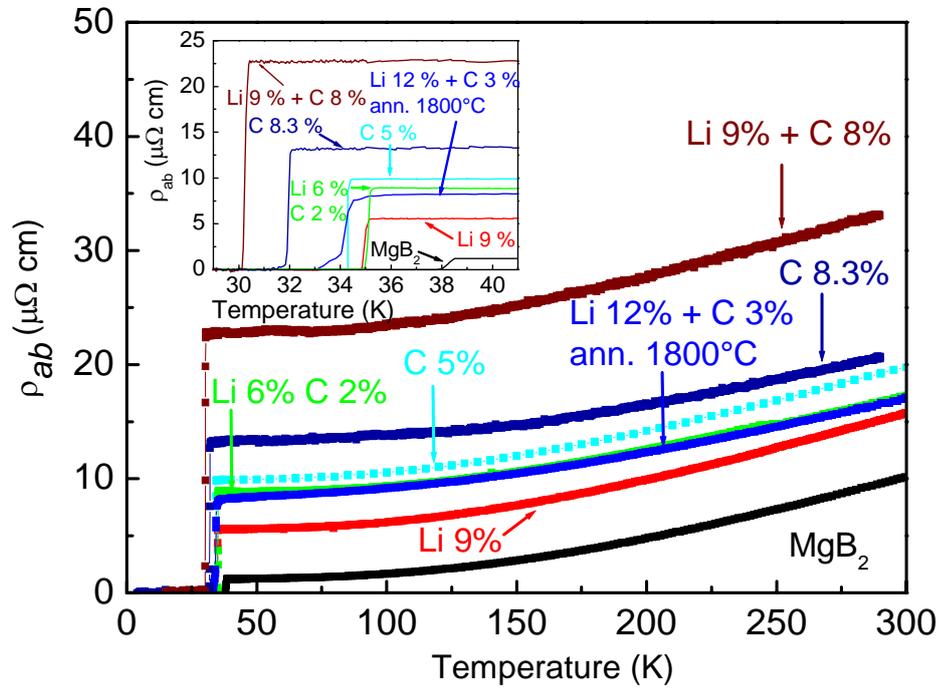

Fig. 15. (Color online) In-plane resistivity as a function of temperature for ~~one~~ single crystals of $Mg_{1-x}Li_xB_2$ with $x=0.09$ and of $Mg_{1-x}Li_x(B_{1-y}C_y)_2$ with $x=0.06$, $y=0.02$; $x=0.12$, $y=0.03$; and $x=0.09$, $y=0.08$. The crystal with $x=0.12$ and $y=0.03$ has been annealed at high pressure at 1800 °C, after completing the growth. Data for pure $MgB_2$ and for $Mg(B_{1-y}C_y)_2$ with $y=0.05$ and $y=0.083$ are presented for comparison. Inset: magnification of the of the resistivity changes in temperature region close to the superconducting transitions.

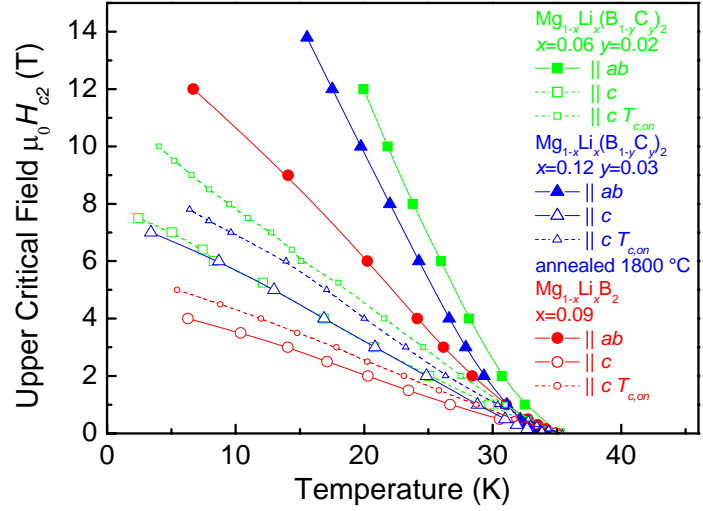

Fig. 16. (Color online) Upper critical fields $H_{c2}$ as a function of temperature for the field parallel to the *ab* plane (closed symbols) and to the *c* axis (open symbols) for Li substituted and for Li-C co-substituted single crystals. The critical field is determined by means of the "zero resistance" definition of $T_c$ (solid lines, large symbols). For the field parallel to the *c* axis, a critical field determined by the "onset" definition of $T_c$ (dashed lines, small symbols) is shown as well.

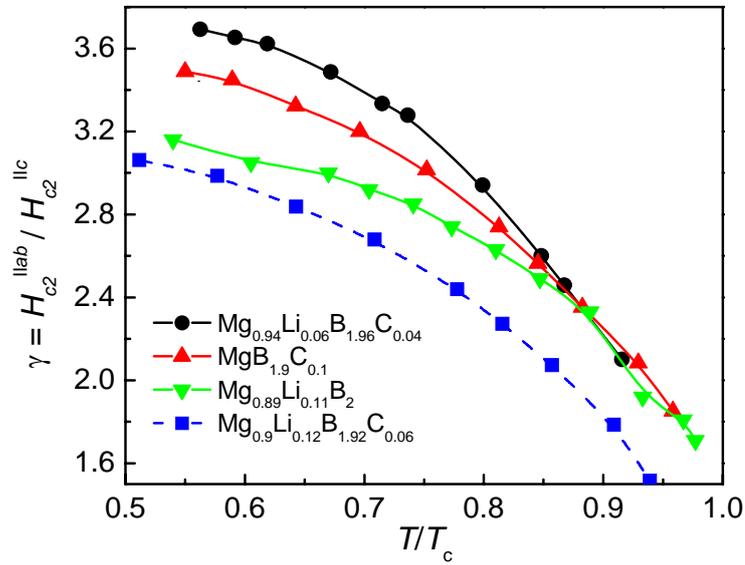

Fig. 17. (Color online) Temperature dependence of upper critical field anisotropy $\gamma$ for crystals with various substitutions and similar $T_c$: Li-C co-substituted ($Mg_{0.94}Li_{0.06}B_{1.96}C_{0.04}$, $T_c$=35.6 K, $Mg_{0.88}Li_{0.12}B_{1.94}C_{0.06}$, $T_c$=34.8 K), Li substituted ($Mg_{0.89}Li_{0.11}B_2$, $T_c$=35.2 K) and C substituted ($MgB_{1.9}C_{0.1}$, $T_c$=34.3 K).